\documentstyle[12pt]{article}
\begin{document}

\begin{center}
{\bf Direct hadron production in ultrarelativistic heavy ion collisions.}

D.Yu.Peressounko, Yu.E.Pokrovsky

{\it Russian Research Center ''Kurchatov Institute'', 123182 Moscow, Russia.}
\end{center}

\begin{center}
{\bf Abstract}
\end{center}

{
Hadrons emitted by the pre-surface layer of quark-gluon plasma
(QGP) before phase transition into hadronic gas are considered as possible
source of direct information about QGP. It is shown that these hadrons
dominate at soft $p_t$ if QGP is created in ultrarelativistic heavy ion
collisions.
}



\newpage

\section{Introduction}

It is commonly believed that hadrons produced in the heavy ion collisions
suffer numerous rescattering in the hot and dense matter before escaping
from it, and thus can not bring out direct information about central part of
the collision. That's why when one studies signatures of quark-gluon plasma
(QGP), which possibly can be created in the heavy ion collisions, one usually
treats hadrons as a particles, carrying information mainly about the
freeze-out
stage of collision. This is true if the size of the region occupied by the
hot matter created in the collision is much greater than the free path length of
the hadron in the hadronic gas. However, even in the collisions of the
heaviest nuclei, the size of the region occupied by hot matter is of the order
of the size of the colliding nuclei and comparable with free path length of
hadron in hadronic gas -- so hadron emission from the central region of the
collision may be significant.

In this paper we consider the emission of hadrons directly from the central hot
region of
the collision. In particular, if QGP is created in the heavy
ion collision, quarks and gluons may fly out from its pre-surface layer,
transform into hadrons on its surface, and some of this hadrons can pass
through surrounding hadronic gas without rescattering. Thus formed
{\it direct} hadrons carry {\it direct} information about QGP surface, in
contrast to usually considered `freeze-out' hadrons originated either due to
freeze-out of hadronic gas or due to decays of the resonances originated due
to freeze-out of hadronic gas.

The effect of meson emission (evaporation) from the QGP surface have been
discussed several times. In the paper \cite{Evaporation1} it was considered
in connection with reducing of plasma pressure and thus influence on the
plasma evolution. It was found that meson evaporation leads to the
significant energy loss and reduction of the outward pressure on the 
QGP surface. Later in \cite{Evaporation2} the same estimation was 
performed, but hadronization of flying out quarks was described in a
frame of the chromoelectric flux tube (CFT) model. It was shown that in 
this model the effect of meson evaporation from the plasma surface is 
small enough to not be important in considering of plasma evolution. It 
should be noted that the CFT model implies that emission takes place 
from the surface of a very dilute system, where the only possibility to 
create a meson is to born quark pair in the tube and thus discolor 
flying out quark. In this paper we consider a somewhat opposite 
situation: we suppose that dense QGP is created and the dominant 
mechanism of discoloring of flying out quarks or gluons is pulling into 
tube of soft quarks and gluons from QGP.

Hadron emission in the heavy ion collisions in the frame of quark-gluon
string model was considered in \cite{FreezeCasc}. It was shown
that in this model final state hadrons are emitted from the whole space-time
volume of the system but not from the thin freeze-out hypersurface. In our
previous work we considered direct hadron emission from the static spherical
volume of the QGP \cite{We}. We showed that in this case direct hadrons
strongly dominate in the soft part of the spectrum. In the present paper we
consider a more realistic model which takes into account both the evolution of the
hot matter and absorption by surrounding hadronic gas.

The layout of the paper is as follows: in section 2 we discuss
our model in detail; in section 3 we apply our model to the S+Au collision at
$200\;A\cdot GeV$ (SpS) and make predictions for Pb+Pb collision at
$3150+3150\;A\cdot GeV$ (LHC); in section 4 we study the sensitivity of
the results to variation of the parameters. The obtained results are
summarized in the conclusion.

\section{Estimation of hadronic emission from the main stages of
evolution of heavy ion collision.}

In our calculations we suppose that hot matter continuously emits {\it direct}
hadrons during all stages of its evolution, from its creation until
freeze-out. The direct hadrons are emitted from any elementary volume in the
depth of hot hadronic gas until freeze-out takes place in this volume, and
from the QGP surface until it transforms into hadronic gas. So in the
collision of the heavy nuclei we have three main hadronic sources: direct
hadrons from the QGP surface, direct hadrons from hadronic gas, and hadrons
originated due to hadronic gas freeze-out (freeze-out hadrons). To calculate
the multiplicity and $p_t$ distribution of the direct hadrons, we evaluate
quark, gluon, and hadron emission rates (i.e. the number of particles
emitted from unit volume per unit time) and integrate them over the
space-time volume occupied by expanding matter. Taking into account
possible rescattering of the flying out direct hadrons, we use the
probability to fly out:
\begin{equation}
\label{Probability}P=\exp \left\{ -\int \frac{dx}{\lambda (x)}\right\} ,
\end{equation}
where integration is done over the path of the hadron in the hot hadronic gas, $
\lambda (x)$ - depending on the local energy density free path length of the
hadron in the hadronic gas. We use the same expression, though with different free path
lengths, for the probability for quarks and gluons to fly out from the QGP
and hadronize on its surface. To incorporate collective phenomena and
energy-momentum conservation in order to find the space-time energy density
distribution and thus estimate absorption by the hadronic gas, we use
a hydrodynamic model of expansion of the hot matter. As a first step we do
not consider emission of heavy mesons and resonances, but restrict ourselves
to the emission of pions. Also in estimates of hydrodynamic expansion we
suppose that hadronic gas consists only of pions. We describe hydrodynamic expansion of
the hot matter using the refined version of Bjorken (scaling)
hydrodynamics including transverse expansion. Hydrodynamic expansion stops
when the energy density in the particular elementary volume becomes as low as the
freeze-out one, and in the further evolution pions in this volume are treated
as free particles.

Below we consider in detail the main points of our calculation: estimates of
the free path lengths of the particles in the hot matter, estimates of the
emission rate of the particles, description of hadronization of quarks or
gluons flying out from the QGP on its surface, and description of the
hydrodynamic expansion.

\subsection{Quark, gluon and pion free path lengths in the medium.}

Let us estimate the free path length of a quark or gluon in the
quark-gluon plasma. Quark or gluon propagation in the
quark-gluon plasma is described by two parameters: color and momentum
relaxation free path lengths. The first is the length on which color of the
quark or gluon changes due to soft gluon exchange with surrounding plasma,
the second is the length on which the momentum of the particle changes significantly
due to more or less hard scattering on gluons and quarks of the plasma.
This two lengths are related as $\lambda _c\approx \alpha _s\lambda _m$.
Our estimates depend only on the momentum relaxation free path lengths. To
estimate them we use the following expression:
\begin{equation}
\label{QGPlam}
\begin{array}{c}
\lambda _i^{-1}(\varepsilon _1)=\sum\limits_j\int \frac 1{2\,\varepsilon _1}
\frac{d^3p_2}{2\varepsilon _2(2\pi )^3}f(\varepsilon _2)\frac{d^3p_3}{
2\varepsilon _3(2\pi )^3}(1\pm f(\varepsilon _3))\frac{d^3p_4}{2\varepsilon
_4(2\pi )^3}(1\pm f(\varepsilon _4)) \\ (2\pi )^4\delta (p_1+p_2-p_3-p)\mid
\!M_{ij}\mid ^2,
\end{array}
\end{equation}
where only two-particle reactions are taken into account, $i,j\in \{q,
\overline{q},g\},$ $\varepsilon ,\,p$ - energy and momentum of the initial
(indexes 1,2) and final (indexes 3,4) partons, $f(\varepsilon )-$
distribution functions (Bose or Fermi), $M_{ij}$ - matrix element of the
reaction. In the case of Boltsman distributions $f(\varepsilon )$
eight integrations in (\ref{QGPlam}) can be performed analytically:

\begin{equation}
\label{QGPlamInt}\lambda _i(\varepsilon )=\left[ \frac 1{16\,\pi ^3}\frac
T{\varepsilon ^2}\sum_j\int\limits_{2\cdot m^2}^\infty s\,\sigma
_{ij}(s)\;\ln \left( 1-\exp \left( -\frac s{4\varepsilon T}\right) \right)
\,ds\right] ^{-1}.
\end{equation}

We use the following cross-sections of the reactions $qq\rightarrow qq,\;q
\overline{q}\rightarrow q\overline{q},\;qq^{\prime }\rightarrow qq^{\prime
},\;qg\rightarrow qg,\;q\overline{q}\rightarrow gg,\;gg\rightarrow gg$
correspondingly ($q$ denotes quark, $g$ -- gluon, $q^{\prime }$ -- quark
with other then $q$ flavor):
\begin{equation}
\label{QGcross}
\begin{array}{l}
\sigma _{qq}=\frac 49
\frac{\pi \alpha _s^2}s\left( 1-2\frac{m^2}s+2\frac s{m^2}-2\frac
s{s-m^2}-\frac 83\ln \left( \frac{s-m^2}{m^2}\right) \right) \\ \sigma _{q
\overline{q}}=\frac{32}{27}\frac{\pi \alpha _s^2}s\left( 1-\frac 94\frac{m^2}
s+\frac 34\frac s{m^2}-\frac 34\frac s{s-m^2}+\frac 34\frac{m^4}{s^2}-\frac
12\frac{m^6}{s^3}-\ln \left( \frac{s-m^2}{m^2}\right) \right) \\ \sigma
_{qq^{\prime }}=\frac 49
\frac{\pi \alpha _s^2}s\left( 1-2\frac{m^2}s+2\frac s{m^2}-2\frac
s{s-m^2}-2\ln \left( \frac{s-m^2}{m^2}\right) \right) \\ \sigma _{qg}=
\frac{31}9\frac{\pi \alpha _s^2}s\left( 1-2\frac{m^2}s+\frac{45}{62}\frac
s{m^2}-\frac{45}{62}\frac s{s-m^2}-\frac{45}{62}\ln \left( \frac{s-m^2}{m^2}
\right) \right) \\ \sigma _{ann}=
\frac{67}9\frac{\pi \alpha _s^2}s\left( 1-\frac{104}{67}\frac{m^2}s-\frac{90
}{67}\frac{m^4}{s^2}+\frac{60}{67}\frac{m^6}{s^3}+\frac{32}{67}\ln \left(
\frac{s-m^2}{m^2}\right) \right) \\ \sigma _{gg}=\frac{123}8\frac{\pi \alpha
_s^2}s\left( 1-\frac{675}{328}\frac{m^2}s+\frac 8{41}\frac s{m^2}-\frac
8{41}\frac s{s-m^2}+\frac{57}{328}\frac{m^4}{s^2}-\frac{19}{164}\frac{m^6}{
s^3}-\frac{57}{164}\ln \left( \frac{s-m^2}{m^2}\right) \right) .
\end{array}
\end{equation}

To incorporate medium effects, we use
screening mass $m^2=4\pi \alpha _sT^2$ instead of current masses of light
quarks and zero mass of gluon (see e.g.\cite{EnergyLoss}). Substituting
cross-sections (\ref{QGcross}) into (\ref{QGPlamInt}), and performing
resting integration numerically with $\alpha _s=0.3$, we obtain quark
and gluon free path lengths in the QGP, see
fig.1. In our calculations we use the following fits to these free path
lengths:

\begin{equation}
\label{QGfit}\,\lambda _q=\frac{0.2}{T^{0.9}},\qquad \lambda _g=\frac{0.45}{
T^{0.9}}.
\end{equation}
where $\lambda $ in $fm$ and $T$ in $GeV$. This result is consistent with
estimates of the energy loss of the quark or gluon in the QGP (see e.g. \cite
{EnergyLoss}). Our estimates are valid if the energy of the particle is
large with respect to the temperature of the plasma, but as a first step we
extrapolate found free path lengths to the energies of the order of
the temperature of the QGP.

To estimate pion free path length in the pionic gas we take advantage of (
\ref{QGPlam}) and perform all possible integrations analytically, assuming
non-zero pion mass. As a result we obtain following formula:

\begin{equation}
\label{PiLamPrec}\lambda _{\pi ^i}(\varepsilon )=\left[ \frac 1{16\,\pi
^3}\frac T{\varepsilon \ p}\sum_j\int\limits_{4\cdot m^2}^\infty \sqrt{
s\,(s-4\,m^2)}\,\sigma _{ij}(s)\;\ln \left( \frac{1-\exp (-a_{+})}{1-\exp
(-a_{-})}\right) \,ds\right] ^{-1}
\end{equation}
where%
$$
a_{\pm }=\frac{\left( s\,\varepsilon -2\,\varepsilon \;m^2\pm p\sqrt{
s^2-4\,s\,m^2}\right) }{2\,m^2T},\quad
$$
$\sigma _{ij}(s)$ -- $\pi ^i\pi ^j$cross-section ( $i,j=\{+,0,-\}$ ), and $m$
- pion mass. All necessary pion-pion cross-sections in the c.m. energy $
\sqrt{s}=0.2-1\;GeV$ (see fig.2) we take from \cite{PionCrossSection}, while
for $\sqrt{s}>1\;GeV$ we take the sum of cross-sections $\sigma
_i=\sum_j\sigma _{ij}$ as constant of order $20\;mb$. Variation of $
\sigma _i$ for $\sqrt{s}>1\;GeV$ leads to slight variations of the slope in
the dependence of free path length on the energy of the pion. Estimations of
the pionic free path length at three different temperatures of pionic 
gas are shown in fig.3.

\subsection{Estimations of the quark, gluon and pion emission rates}

In this subsection we estimate emission rates of the quark and gluon from
quark-gluon plasma and pions from pionic gas. In this calculation we assume
that the particle is emitted at the moment of its last scattering on other
one or at the moment when the particle is born. Thus we do not take into
account a few particles which born before we begin consideration of the
evolution of the system and pass through the system without rescattering.

In our estimates of quark and gluon emission rate we take into account only
two-particle reactions like $qq\rightarrow qq,\;q\overline{q}\rightarrow q
\overline{q},\;qq^{\prime }\rightarrow qq^{\prime },\;qg\rightarrow qg,\;q
\overline{q}\rightarrow gg,\;gg\rightarrow gg$ -- the same as those, taken
for the free path lengths evaluation. Thus the emission rate of the parton $i$
from quark-gluon plasma can be written as
\begin{equation}
\label{rates}
\begin{array}{c}
\varepsilon
\frac{d^7R_i}{d^3p\,d^4x}=\sum\limits_j\int \frac{d^3p_1}{2\varepsilon
_1(2\pi )^3}f(\varepsilon _1)\frac{d^3p_2}{2\varepsilon _2(2\pi )^3}
f(\varepsilon _2)\frac{d^3p_3}{2\varepsilon _3(2\pi )^3}(1\pm f(\varepsilon
_3))\times \\ (1\pm f(\varepsilon ))\frac{(2\pi )^4\delta (p_1+p_2-p_3-p)}{
2(2\pi )^3}\mid \!M_{ij}\mid ^2,
\end{array}
\end{equation}
where the sum is over all possible two-particle reactions, \ $p_1,\,p_2$ --
momentum of the initial particles, $p_3,\,p$ -- momentum of the final
particles, $f(\varepsilon )$ -- Fermi or Bose distributions for quarks and
gluons respectively, $(1\pm f(\varepsilon ))$ -- Fermi blocking or Bose
enhancement for final state quarks and gluons respectively, and $M_{ij}\,$
-- matrix element for the process of the $i$ on $j$ scattering.

This formula is in agreement with estimates of quark and gluon emission from
the unit of surface per unit of time of semi-infinite QGP with constant
temperature. On one hand this is just thermal black body radiation and
described by the formula:
$$
\varepsilon \frac{d^6R^{thermo}}{d^3p\,d^2S\,\,dt}=\frac{d\cdot \varepsilon
}{(\exp (\varepsilon /T)\pm 1)}
$$
where $d$ -- degeneracy ($d=16$ for gluons, $d=24$ for $u,d$ quarks). On
the other hand we can evaluate the same quantity from the microscopic
approach -- via emission rate and momentum relaxation free path length:
\begin{equation}
\label{link}\varepsilon \frac{d^6R_i^{micro}}{d^3p\,d^2S\,dt}=2\pi \int
dx\int \sin (\theta )\,d\theta \;\exp \left( -\frac x{\lambda (\varepsilon
)\cdot \cos (\theta )}\right) \cdot \varepsilon \frac{d^7R_i}{d^3p\,d^4x}.
\end{equation}

Equation (\ref{link}) relates emission rate with free path length as:

\begin{equation}
\label{NewRate}\varepsilon \frac{d^7R_i}{d^3p\,d^4x}=\frac{d_i}{\lambda
_i(\varepsilon )}\frac \varepsilon {(\exp (\varepsilon /T)\pm 1)}.
\end{equation}

In our calculations we use (\ref{NewRate}) instead of separate evaluation of
free path length and emission rate.

To take into account the collective velocity of the emitting volume, we use
the invariance of $\varepsilon \,d^3\,R\,/\,d^3p$ with respect to Lorenz boosts.
So the right side of the (\ref{rates}) is invariant too. In the rest frame
of the volume we have $\varepsilon \,d^3R\,/\,d^3p=f(\varepsilon ) $, then
in the laboratory frame, where volume moves with 4-velocity $u$, we have $
\varepsilon \,d^3\,\!R\,/\,d^3p=f(p\cdot u),$ where $p$ is $4$-momentum of
the particle.

\subsection{Hadronization of the quarks and gluons on the plasma surface.}

To describe hadronization of a quark or gluon crossing the QGP surface, we
assume that the flying out parton deforms the plasma surface by creation of a
string-like tube, which breaks if the colorless state is formed in the tube and
one final hadron is born. The energy of the final hadron is in the range $
m_h<\varepsilon <\varepsilon _p$ with constant probability for each
value. So $p_t$ distribution of the partons $d^3R^{part}/dy\,d^2p_t,$
determines $p_t$ distribution of the final hadrons:
\begin{equation}
\label{R3-hadr} \frac{d^3R^{hadr}}{dy\,d^2p_t}=\frac 1\varepsilon
\int\limits_\varepsilon ^\infty d\omega \cdot \frac{d^3R^{part}}{
dy\,d^2p_t},
\end{equation}
where it is assumed that one flying out parton transforms into one final
hadron. It can be easily seen that this formula describes the decay of the
particle given only one restriction: the energy of the initial particle
must be greater than energy of the final one. In this paper we take into
account decays into pions only, inclusion of the resonances is in progress.

We stress that the hadronization of the quark or gluon in this process
differs from hadronization in the $e^{+}e^{-}$ annihilation. If QGP
is formed then a quark or gluon flying out from its surface may 'pull'
soft quarks and gluons from the plasma into the string.
As a result of this discoloring the string is broken and final state hadron
is formed.
In the $e^{+}e^{-}$ annihilation the only possibility to create a final
hadron is to
born $q\overline{q}$ pair in the string. So the physics of these two
processes is different and applicability of the string model as it is used in
the description of $e^{+}e^{-}$ annihilation to the description of the quark and
gluon hadronization on the QGP surface is questionable.

A few words should be added about formation of the direct pions emitted
from QGP. The pion is a rather extensive and complicated object,
and it takes some time to form its structure from the piece of the
string flying out from QGP. This formation of the structure takes at least
the time needed for light to cross the pion, i.e. approximately $0.6\;fm/c$ in
the rest frame. During this formation time the pion does not interact with
surrounding hadronic gas as a real pion, but rather as a pair of quarks.
Corresponding quark-quark cross-sections $\sigma _{qq}\sim 2-10\;mb$ are
approximately 10 times smaller then $\sigma _{\pi \pi }\sim 20-130\;mb$, and
in our calculations we accept that during formation time of order $\tau
_0=0.6\;fm/c$ in the rest frame after emission from the plasma surface the
emitted pion does not interact with hadronic gas.

\subsection{Hydrodynamic description of the evolution.}

In order to take into account absorption of the direct hadrons in hot
expanding hadronic gas surrounding QGP and estimate time of life of the QGP
we use a hydrodynamic approach. This is just an expression of local energy and
momentum conservation:
\begin{equation}
\label{dT}\partial _\mu T^{\mu \nu }=0,
\end{equation}
where $T^{\mu \nu }$ is the energy-momentum tensor:
\begin{equation}
\label{T=}T^{\mu \nu }(x)=\left( \varepsilon (x)+p(x)\right) \cdot u^\mu
u^\nu -g^{\mu \nu }\cdot p(x).
\end{equation}

Here $\varepsilon (x)$ and $p(x)$ -- local energy density and pressure
respectively, $u^\mu $ -- four-velocity of the medium and $g^{\mu \nu }$
metric tensor ($g=diag(1,-1,-1,-1)$). Having in mind applications of our
approach to predictions for LHC energy, we introduce some simplifications
into initial conditions and equations suitable for high energy case: we
suppose that Bjorken (scaling) hydrodynamics \cite{Bjorken} is applicable.
Simplifications, proposed by Bjorken, are based on the fact that in
midrapidity interval all quantities are independent of rapidity. So
hydrodynamic equations must be invariant with respect to the Lorenz boosts
along collision axis. That is four-velocity can be parametrized in the form
$$
u^\mu =\gamma \,(\frac t\tau ,\overrightarrow{v},\frac z\tau ),
$$
where $\tau =\sqrt{t^2-z^2},\gamma =\left( 1-v^2\right) ^{-\frac 12}$,$\;v$
-- transverse velocity. With these assumptions the hydrodynamic equations
in the midrapidity region ($z=0$) can be rewritten as
\begin{equation}
\label{part}
\begin{array}{l}
{\partial _\tau \left( \gamma ^2w(r,\tau )\right) +\partial _r\left( \gamma
^2vw(r,\tau )\right) + \gamma ^2w(r,\tau )\tau^{-1} + \gamma
^2vw(r,\tau )r^{-1}-\partial _\tau p(r,\tau )=0} \\ {\partial _\tau \left(
\gamma ^2vw(r,\tau )\right) +\partial _r\left( \gamma ^2v^2w(r,\tau )\right)
+\gamma ^2vw(r,\tau )\tau^{-1} +\gamma ^2v^2w(r,\tau )r^{-1}+\partial
_rp(r,\tau )=0},
\end{array}
\end{equation}
where $w(r,\tau )=\varepsilon (r,\tau )+p(r,\tau )$ -- entalpy density.

We solve equation (\ref{part}) numerically using the MacCormack technique. This
is explicit predictor-corrector Euler algorithm, which allows second order accuracy
on a spatial and time grid step. To rewrite (\ref{part}) in
terms of final differences it is convenient to introduce the following
variables:
$$
U=\left(
\begin{array}{c}
{\gamma ^2w-p} \\ {\gamma ^2vw}
\end{array}
\right) ;\quad F=\left(
\begin{array}{c}
{\gamma ^2vw} \\ {\gamma ^2v^2w+p}
\end{array}
\right) ;\quad H=\left(
\begin{array}{c}
{\gamma ^2w\left( \tau^{-1} + v r^{-1}\right) } \\ {\gamma ^2vw\left(
\tau^{-1} + v r^{-1}\right) }
\end{array}
\right)
$$

In these variables equation (\ref{part}) can rewritten in a discrete form
(subscript correspond to the spatial sell number, superscript correspond to
the time level):
\begin{equation}
\label{McCorm}
\begin{array}{l}
{\overline{U}_i^j=U_i^j-\Delta\tau \left( \left( F_{i+1}^j-F_i^j\right)
/\Delta r+H_i^j\right) +D_i^j} \\ {U_i^{j+1}=\frac 12\left[ \left( U_i^j+
\overline{U}_i^j\right) -\Delta \tau\left( \left( \overline{F}_i^j-
\overline{F}_{i-1}^j\right)
/\Delta r+\overline{H}_i^j\right) \right] +\overline{D}_i^j}
\end{array}
\end{equation}
where $\overline{F}_i^j,\overline{H}_i^j,\overline{D}_i^j$ are evaluated
using $\overline{w}_i^j,\overline{v}_i^j$ taken from $\overline{U}_i^j$, and
$D_i^j$ -- term, corresponding to the smoothing, proposed by Zmakin and
Fursenko \cite{Smoothing}. This term conserves global energy and momentum
and allows working with various hydrodynamic disturbances such as shock waves
and first order phase transition. It can be written as
\begin{equation}
\label{Smoothing}D_i^j=\left\{
\begin{array}{l}
{\varphi _i^j,\quad \left( \varphi _{i-1}^j\cdot \varphi _{i-2}^j<0\right)
\cup \left( \varphi _i^j\cdot \varphi _{i-1}^j<0\right) } \\ {0,\quad
\;\left( \varphi _{i-1}^j\cdot \varphi _{i-2}^j>0\right) \cap \left( \varphi
_i^j\cdot \varphi _{i-1}^j>0\right) }
\end{array}
\right.
\end{equation}
where $\varphi _i^j=Q\cdot \left( U_{i+1}^j-U_i^j\right) ,$ and $Q$ --
parameter of smoothing (the best results are obtained with
$Q\approx 0.1-0.15$).

Hydrodynamic equations must be accomplished by equations of state. In our
calculations we used the simplest one, including first order phase transition
at the temperature $T_c$ and freeze-out of the pionic gas at the
temperature $T_{freeze}$:
\begin{equation}
\label{eq-st}P(\varepsilon )=\left\{
\begin{array}{cc}
\frac \varepsilon 3-\frac 43B, & \varepsilon \geq \varepsilon _{qgp} \\
P_0, & \varepsilon _{qgp}\geq \varepsilon \geq \varepsilon _{had} \\
\frac \varepsilon 3, & \varepsilon _{had}\geq \varepsilon \geq \varepsilon
_{freeze} \\
0, & \varepsilon _{freeze}\geq \varepsilon
\end{array}
\right.
\end{equation}
where $B$ -- energy density of perturbative vacuum of QCD (bag constant),
$\varepsilon _{qgp}$ -- energy density of the QGP
before phase transition, $\varepsilon _{had}$ -- energy density of the
pionic gas after phase transition and $\varepsilon _{freeze}$ -- energy
density of the pionic gas at freeze-out:%
$$
\varepsilon _{qgp}=d_{qgp}\frac{\pi ^2}{30}T_c^3+B,\qquad \varepsilon
_{had}=d_{had}\frac{\pi ^2}{30}T_c^3,\qquad \varepsilon _{freeze}=d_{had}
\frac{\pi ^2}{30}T_{freeze}^3
$$
where $d_{qgp}=37$ -- degeneracy of the QGP, $d_{had}=3$ -- degeneracy of
the pionic gas.

It is well known that the hydrodynamic description assumes local thermal
equilibrium, which surely can not be reached in the very beginning of the
collision when the colliding nuclei penetrate each other. So to properly describe
a collision we have to either describe the pre-equilibrium dynamics
(i.g. three-fluid hydrodynamics \cite{ThreeFluid}) or introduce some
parameters such as initial time (initial volume) and initial spatial
energy density and velocity distributions. In this paper we use the latter
approach: we introduce these parameters into the scheme and adjust them to
reproduce measured pion multiplicity and $p_t$ spectrum. In our estimates we
use an initial spatial energy density distribution proportional to the
integral of the nuclear energy density along the beam axis:
\begin{equation}
\label{Nuclear}\varepsilon _{in}(r)=\varepsilon _0\int\limits_{-\infty
}^\infty \left[ 1+\exp \left( \frac{(\sqrt{r^2+l^2}-R_{in})}w\right) \right]
^{-1}dl,
\end{equation}
where $\varepsilon _0=d_{qgp}\frac{\pi ^2}{30}T_{in}^3+B$ if we assume QGP
formation, and $\varepsilon _0=d_{had}\frac{\pi ^2}{30}T_{in}^3$ in the case
of pure hadronic gas. $R_{in}$ and $w$ are taken from known nuclear density
distribution, e.g. for nuclei $S^{32}$ we have $R_{in}=3.5\;fm,\,w=0.65\;fm$.
For initial radial velocity distribution, we take $v_{in}(r)=0$.

\section{Obtained results.}
\subsection{Predictions for S+Au collision at 200 $A\cdot GeV$.}

Now let us apply our model to the S+Au collisions at 200 $A\cdot GeV$, which
were studied experimentally. We
consider two possible cases of evolution: with and without QGP creation. In
both cases we adjust the free parameters of the model to reproduce measured
(\cite{DataWA80}) pion multiplicity and $p_t$ distribution at midrapidity. In
the case of QGP creation, the experimental data are better described by the
following set of parameters:
$$
T_{in}=180\;Mev,\;\tau
_{in}=1.3\;fm/c,\;R_{in}=3.5\;fm,\;T_c=150\;MeV,\;T_{freeze}=140\;MeV.
$$
In the case of pure hadronic gas the better description is obtained 
with another set:  
$$ T_{in}=180\;MeV,\;\tau 
_{in}=4.3\;fm/c,\;R_{in}=7\;fm,\;T_{freeze}=140\;MeV.  
$$

Estimation of $p_t$ distributions of $\pi ^0$ at midrapidity ( $y\approx 3$)
for the different pionic sources in this reaction with assumption of QGP
creation is shown in fig.4. The dotted line corresponds to the direct
pions from QGP surface, the dashed line corresponds to the direct 
pions from the depth of pionic gas and the solid line corresponds 
to the freeze-out pions. Direct pions from QGP surface dominate in 
the $p_t$ range $0-0.5\;GeV/c$ while at large $p_t$ freeze-out 
pions dominate.

Naively one
could expect the inverse situation: emissions from the hotter phase dominate in the
hard part of the spectrum while emissions from the cooler phase dominate in the
soft one. However, if we take into account the collective velocity of
pionic gas in hydrodynamic expansion, we find that the effective
temperature of the spectrum of freeze-out pions $T^{\prime }=T\sqrt{
(1+u)/(1-u)}$ ($u$ -- collective velocity of the volume, $T$ - temperature
of pionic gas in the rest frame) for reasonable collective velocities
$u>0.3$ is larger than even initial temperature. Anywhere, direct pions
strongly contribute to total multiplicity: at midrapidity multiplicity of
the direct pions is $dN^{dir}/dy=70$ while multiplicity of freeze-out ones is
$dN^{freeze}/dy=50$. So even for collisions of not too heavy nuclei at
relatively low energies, when QGP (if created) lives a rather short time,
direct pion contribution to total pionic yield is significant.

If we do not assume QGP creation, and consider evolution of pure pionic
gas, we obtain $p_t$ distributions of $\pi^0$ at midrapidity ($y\approx 3$),
shown in fig. 5. Dashed line corresponds to the direct pions from pionic
gas, solid line corresponds to the freeze-out pions. In this case of
evolution direct pions do not dominate anywhere and the main 
contribution comes from freeze-out pions: multiplicity at midrapidity 
of the direct pions is $dN^{dir}/dy=8$ while this value for freeze-out 
pions is $dN^{freeze}/dy=110$.

The comparison of total $p_t$ distributions of $\pi ^0$ at midrapidity,
calculated assuming two possible cases of evolution with experimental data
is shown in fig.6. The dotted line corresponds to the estimate assuming
QGP formation, the solid line corresponds to the estimate considering pure
pionic gas evolution. Circles correspond to $p_t$ distribution of $\pi 
^0$, obtained by WA80 collaboration for S+Au at 200 $A\cdot GeV$ 
\cite{DataWA80}.  The results of this calculation, assuming pure 
pionic gas evolution, better describe experimental results, except a 
region at small $p_t$. This exceeding of the experimental distribution 
can be explained by decays of resonances ($ N,\Delta $, etc.). In the 
case of QGP creation it is impossible to describe $ p_t$ distribution 
as well as in the case of pure pionic gas, as long as we use 
reasonable values for parameters of the model.

Direct pions may carry out a significant part of the energy from the 
central hot region of the collision and thus strongly affect the 
evolution of the system. To compare evolution of the hot matter in the 
S+Au collision at 200 $A\cdot GeV$ with direct pion production and 
without it, we plot contour curves for constant energy density for 
these two cases -- see fig.7. The left plot corresponds to the 
evolution with direct pion emission, the right plot corresponds to 
the evolution without direct pion emission. Initially the system is 
in the QGP state (black region), then it transforms into mixed phase, 
expands longitudinally and transversely and thus cools, until pure 
pionic gas is formed (gray region). We see that if we take into account 
direct pion emission, then time of life decreases approximately 1.3 
times in comparison to the case of evolution without direct pion 
emission.  Contour curves for constant velocity from $0.1$ with step 
$0.1$ for this two cases are shown in fig.8. Our calculation stops when 
all pionic gas freezes-out, so in fig. 8 contour curves are shown 
only for $t<12\;fm/c$ for the first case and for $t<15\;fm/c$ for the 
second case. If we compare fig.7 and fig.8, we see that matter is 
accelerated mainly during expansion of pionic gas, mixed phase as a 
phase with $dP/dr=0$, where $P$ - pressure, does not contribute into 
this acceleration. In the first case the velocity obtained by pionic 
gas is lower than the one obtained in the second case -- because of 
longer life time and thus longer acceleration.

Therefore if QGP is formed in this collision, direct pions are 
abundantly emitted from its surface and strongly contribute to the 
total pionic multiplicity. Direct pions carry out a significant 
part of the energy from the central part of the collision and thus 
decrease life time of the system.

\subsection{Prediction for Pb+Pb collision at $3150+3150 \;A\cdot GeV$.}

In the Pb+Pb collisions at LHC energy space-time volume occupied by QGP is
expected to be much larger then one at the SpS energies, and thus direct
pion emission is expected to be much more intense. In this section we
estimate contribution of the direct pions to the total pionic yield in
Pb+Pb collision at LHC energy. We use the following initial conditions,
corresponding to total multiplicity $dN/dy\sim 4000$, predicted by
Monte-Carlo event generators:
$$
T_{in}=350\;Mev,\;\tau _{in}=1\;fm/c,\;R_{in}=6.5\;fm,
$$

and transitions and freeze-out temperatures, used in description of S+Au data:
$$
T_c=150\;MeV,\;T_{freeze}=140\;MeV.
$$

The result of this evaluation is shown in fig.9. The dotted line corresponds to
the direct pions from QGP surface, the dashed line corresponds to the direct
pions from pionic gas and the solid line corresponds to the freeze-out
pions. Direct pions from QGP surface dominate up to an order of magnitude
in the soft part of the distribution $p_t<0.7\;GeV$ while freeze-out pions dominate in the
hard part. Direct pions from pionic gas do not dominate anywhere. We find
that multiplicities per unit rapidity at $y\approx 0$ of direct and
freeze-out pions are $dN^{dir}/dy=3000,\;dN^{freeze}/dy=1200$ respectively.

Comparison of space-time evolutions of the hot matter created in the
collision with and without direct pion emission can be made using contour
curves of constant energy density shown in fig.10. The black region
corresponds to the pure QGP phase, the white region corresponds to the mixed
phase and the gray region corresponds to the pure pionic gas. If we do
not take into account direct pion emission, then the collective
flow becomes more significant (bumps on the contour curves on the right
plot), otherwise cooling takes place mainly due to direct pion emission
(smoother curves on the left plot).

Thus the direct pions are abundantly emitted by surface of QGP if it is created
in the Pb+Pb collision at LHC energy, such that $\sim 70\%$ of pions
produced in this collision are direct pions flying out from the QGP surface
without further rescattering in surrounding pionic gas and carrying out
direct information about QGP. Also the direct
pions carry out a significant part of energy and influence the evolution of
the system: the system cools more due to direct pion emission rather then
hydrodynamic expansion.

\section{Stability to variation of parameters.}

We see that direct pions emitted from QGP surface dominate at small $p_t$
at SpS and LHC energies. This effect is not large enough to be
ensured that it will not disappear for any reasonable choice of the model
parameters. Indeed, we use several parameters (such as freeze-out
temperature $T_{freeze}$, phase transition temperature $T_c$, quark, gluon,
and pion free path lengths, initial energy density and velocity
distributions, initial time) which can not be evaluated exactly; instead, we
obtain their value from more or less reliable estimates and find a
relatively wide range of their possible values. So in this section
we explore sensitivity of our results to variations of values of the
parameters in reasonable limits.

The freeze-out temperature $T_{freeze}$ can be varied in the range
$100-140\;MeV$. Varying $T_{freeze}$ we also have to vary initial temperature
$T_{in}$ to reproduce experimental $p_t$ distribution. Results of
corresponding estimates
(assuming QGP creation) are shown in fig.11. Lines marked by stars
correspond to $T_{in}=150\;MeV,T_{freeze}=100\;MeV$ and lines marked by
circles correspond to $T_{in}=180\;MeV,T_{freeze}=150\;MeV$. Here, and until
the end of the section, dotted, dashed, and solid lines correspond respectively to pions
emitted directly from QGP, pions from pionic gas, and freeze-out pions. We see
that the distribution of direct pions from QGP remains unaffected, the
yield of direct pions from pionic gas slightly increases and $p_t$
the distribution of freeze-out pions becomes harder. Anyway direct pions from
QGP dominate at soft $p_t$ for all tested $T_{freeze}$.

Next, consider the transition temperature $T_c$ (and related to
it bag constant $B$ and external pressure on QGP). Results of estimates
using two different values of $T_c$ are shown in fig.12. Lines marked
by stars correspond to $T_c=150\;MeV$, lines marked by circles correspond
to $T_c=180\;MeV$. Increasing $T_c$ leads to decreasing the life time of
the system (especially of the pure QGP and mixed phase) and thus to decreasing
the multiplicity of direct pions from QGP. The longer evolution of the
pure pionic gas (larger difference $T_c-T_{freeze}$) leads to increasing the
multiplicity of direct pions from pionic gas and a slight increasing of
the slope of freeze-out pions due to larger collective velocity.

A sensitivity of our results to variation of the quark and
gluon free path lengths in QGP is shown in fig.13. Lines marked by circles
correspond to evaluation using the values calculated in section 2.1 for
quark and gluon free path lengths (\ref{QGfit}). Lines marked by stars
correspond to evaluation with free path lengths twice larger then
(\ref{QGfit}). We see that in the case of
larger free path length, and thus lower emission rate of quarks and gluons
(see (\ref{NewRate})), the multiplicity of direct pions from QGP decreases, and
(as a result of energy conservation) the multiplicity of direct pions from
pionic gas and freeze-out pions increases.

To complete investigation of sensitivity of our results to the choosing of
set of parameters, let us vary the free path length of pion in the pion gas ($
\lambda _\pi $). Results of calculations using two pion free path lengths:
estimated in section 2.1 (lines marked by circles) and twice larger
(lines marked by stars), are shown in fig.14. Distribution of direct pions
from QGP does not change significantly, so we can conclude that the depth
of surrounding pionic gas is not large and absorption is low. In accordance
with (\ref{NewRate}) emission rate of pions from pionic gas decreases with
increasing of $\lambda _\pi $ and thus the multiplicity of direct pions from
pionic gas decreases as well.

So the domination of direct pions in the soft region is not an effect of
the accidental choice of a set of parameters of the model, but a real effect, stable
with respect to variation of the parameters.

\section{Conclusions}

We considered effect of emission of direct hadrons from the central region of
the central heavy ion collision: emission of quarks and gluons from depth of
QGP, their hadronization on its surface and partial flying out through
surrounding hadronic gas. As a first step we estimate only emission of pions. 

Estimates performed for S+Au collision at $200\;A\cdot GeV$ show that if
QGP is created then direct pions from the QGP surface dominate in the soft 
region $p_t<0.5\;GeV/c$.

In the case of pure pionic gas we find that direct pions from the depth
of the gas do not dominate anywhere.

Estimation performed for Pb+Pb collision at LHC energy ($dN/dy=4200 $)
shows that direct pion emission becomes more important at
this energy level. We find that multiplicity of direct pions $dN^{dir}/dy=3000$
while multiplicity of freeze-out pions $dN^{freeze}/dy=1200$ -- i.e.
approximately 70\% of final pions come from the QGP surface.

The direct pion emission decreases lifetime of the S+Au system at 200
$A\cdot GeV$ and Pb+Pb at $3150+3150 \;A\cdot GeV$ approximately by 1.3 times.

This result is stable with respect to variation of the model parameters
in a wide reasonable range.

\newpage

\section*{Figure caption}

Fig.1. Free path length of a quark (q) and gluon (g) in the QGP for two
temperatures of QGP and fit to them, used in our analysis.

Fig.2. Pion-pion cross-sections for c.m. energy $\sqrt{s}<1\;GeV$
\cite{PionCrossSection}.

Fig.3. Free path length of a pion in pionic gas for three different
temperatures of the pionic gas.

Fig.4. Differential multiplicity of direct $\pi^0$ from QGP (dotted line),
direct $\pi^0$ from pionic gas (dashed line) and freeze-out $\pi^0$ (solid
line) in the S+Au collision at $200\;A\cdot GeV$ in the case of QGP creation.

Fig.5. The same as fig.4, but for the case of pure pionic gas.

Fig.6. Total differential multiplicities of $\pi^0$ for S+Au collision at
midrapidity ($y\approx 3$) for S+Au collision at $200\;A\cdot GeV$. Circles
-- distribution measured by WA80 collaboration, dotted line -- result of
estimate with QGP creation, solid line -- result of estimate with
pure pionic gas evolution.

Fig.7. Constant energy density contour curves for S+Au collision at $
200\;A\cdot GeV$ for two kinds of evolutions: with and without direct pion
emission. The black region corresponds to the pure
QGP phase, the grey region to the pure pionic gas phase.

Fig.8. Constant velocity contour curves from 0.1 with step 0.1 for S+Au
collision at $200\;A\cdot GeV$ for two kinds of evolutions: with and without
direct pion emission.

Fig.9. The same as fig.4, but for Pb+Pb collision at LHC energy.

Fig.10. The same as fig.7, but for Pb+Pb collision at LHC energy.

Fig.11. Sensitivity of predictions to variation of freeze-out ($T_{freeze}$) and
initial ($T_{in}$) temperatures.

Fig.12. Sensitivity of predictions to variation of transition ($T_c$) temperature.

Fig.13. Sensitivity of predictions to variation of free path lengths of quarks
($\lambda_q$) and gluons ($\lambda_g$) in QGP.

Fig.14. Sensitivity of predictions to variation of free path length of pions
in the pionic gas ($\lambda_{\pi}$).


\begin{thebibliography}{99}
\bibitem{Evaporation1}  M. Danos, J.Rafelski, Phys. Rev. {\bf D 27} (1983)
671.

\bibitem{Evaporation2}  B. Banerjee, N. K. Glendenning, T. Matsui, Phys.
Let. {\bf 127B} (1983) 453.

\bibitem{FreezeCasc}  L.V. Bravina et al., Phys. Let. {\bf B 354} (1995), 196.

\bibitem{We}  D. Yu. Peressounko, Yu. E. Pokrovsky, T. V. Mukhanova, Phys.
of Atomic Nuclei, {\bf 58}, (1995) 1450.

\bibitem{PionCrossSection}  O. O. Patarakin, V. N. Tikhonov, private
communications, the basis of the analysis can be found in O. O. Patarakin et
al., Nucl. Phys. {\bf A 598} (1996) 335.

\bibitem{pion-crossec-dens}  J. L. Goity, H. Leutwyller, Phys. Let. {\bf B 28
} (1989), 517.{\bf \ }

\bibitem{EnergyLoss}  Xin-Nian Wang, Miclos Gyulassy, Michael Pl\"umer,
Phys. Rev. {\bf D 51 }(1995), 3436.

\bibitem{PhotQGP}  J. Kapusta, P. Lichard, D. Seibert, Phys. Rev. {\bf D 44}
(1991), 2774.

\bibitem{PhotRho}  Li Xiong, E. Shuryak, G. E. Brown, Phys. Rev. {\bf D 46}
(1992) 3798.

\bibitem{PhotAx}  K. Haglin, Phys. Rev. {\bf C 50} (1994) 1688.

\bibitem{Landau}  L. D. Landau, Izv. Acad. Nauk XVII Ser. Phys. (1953), 51.

\bibitem{Bjorken}  J. D. Bjorken, Phys. Rev. {\bf D 27} (1983), 140.

\bibitem{Smoothing}  Zmakin, Fursenko, Sov. Journal Vych. Math. and Math.
Phys. v.20,4,p.1021.

\bibitem{ThreeFluid}  L. P. Csernai et al., Phys. Rev. {\bf C26 }(1982) 149.

\bibitem{DataWA80}  R. Albrecht et al., Phys. Lett. {\bf B 361} (1995), 14.
\end{thebibliography}
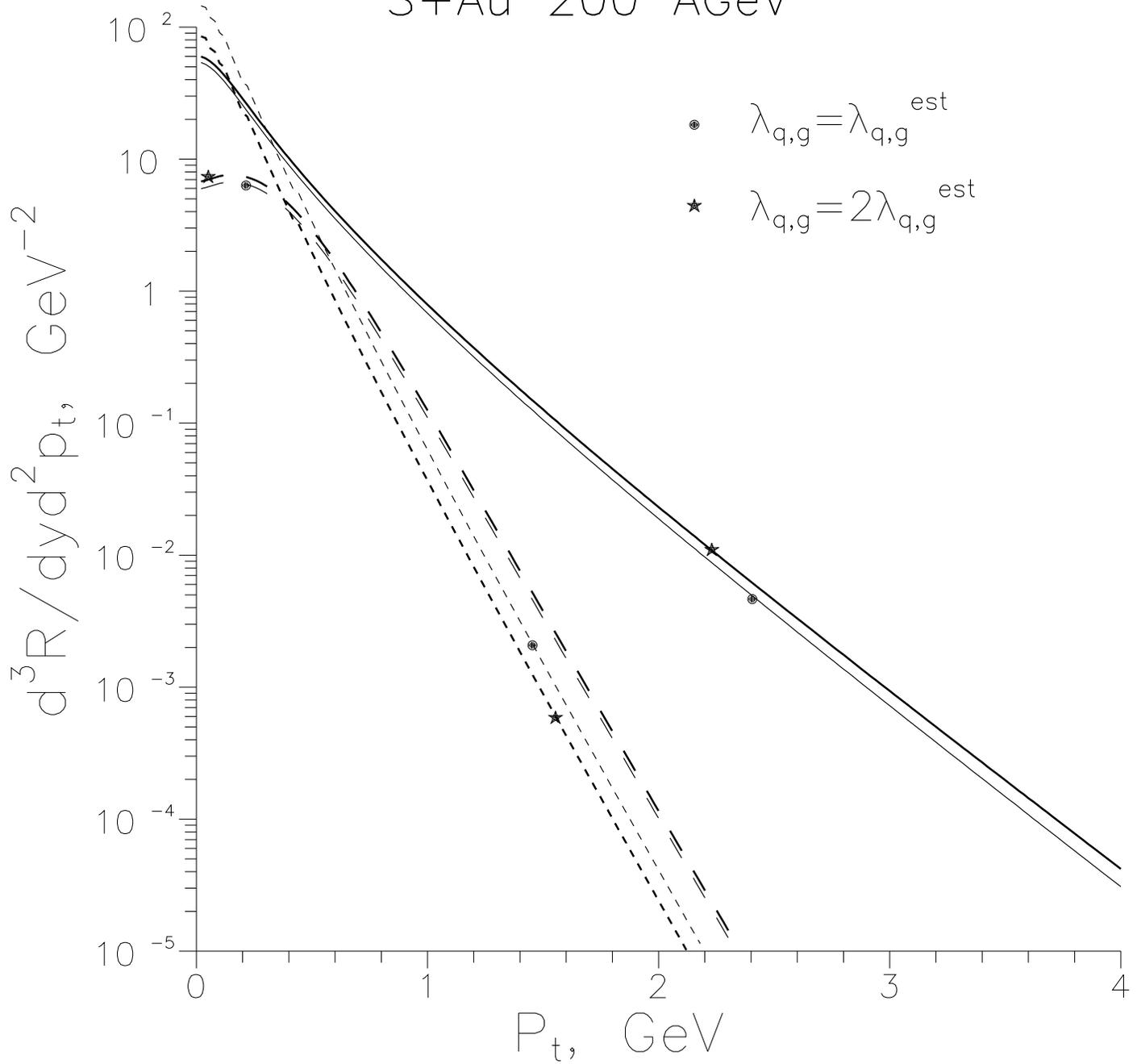
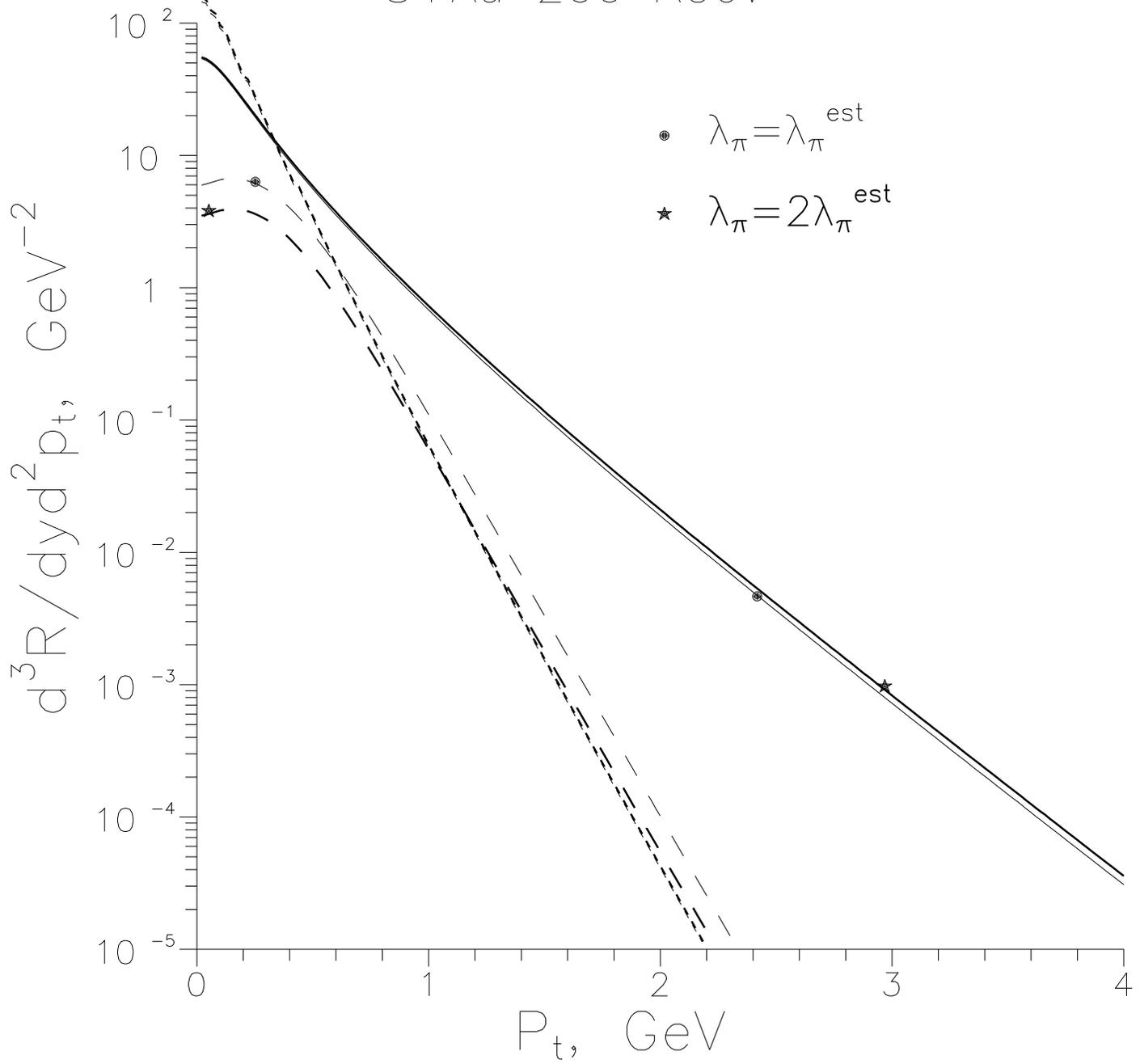
\end{document}